\documentclass[aps,preprint,prd,showpacs,nofootinbib]{revtex4}

\usepackage{amsmath,amssymb}
\usepackage{graphicx,subfigure}
\usepackage{color,multirow}
\usepackage[colorlinks,linkcolor=magenta,anchorcolor=cyan,citecolor=blue,plainpages=false]{hyperref}

\hypersetup{colorlinks=true,
    breaklinks=true,
    pdfstartview=Fit,
    linkcolor=blue,
    citecolor=blue,
    urlcolor=blue}

\def\be{\begin{equation}}
    \def\ee{\end{equation}}
\def\ba{\begin{eqnarray}}
    \def\ea{\end{eqnarray}}

\begin{document}
\title{Impact of a negative cosmological constant on the reconstruction of dark energy in light of DESI BAO data}

    \author{Hao Wang$^{1,2} $\footnote{\href{wanghao187@mails.ucas.ac.cn}{wanghao187@mails.ucas.ac.cn}}}
    \author{Yun-Song Piao$^{1,2,3,4} $ \footnote{\href{yspiao@ucas.ac.cn}{yspiao@ucas.ac.cn}}}

    \affiliation{$^1$ School of Fundamental Physics and Mathematical
        Sciences, Hangzhou Institute for Advanced Study, UCAS, Hangzhou
        310024, China}

    \affiliation{$^2$ School of Physical Sciences, University of
        Chinese Academy of Sciences, Beijing 100049, China}

    \affiliation{$^3$ International Center for Theoretical Physics
        Asia-Pacific, Beijing/Hangzhou, China}

    \affiliation{$^4$ Institute of Theoretical Physics, Chinese
        Academy of Sciences, P.O. Box 2735, Beijing 100190, China}

    \begin{abstract}
An anti-de Sitter vacuum, corresponding to a negative cosmological
constant (NCC), might coexist with one evolving positive dark
energy component at low redshift and is hinted by the latest DESI
observations. In this paper, we use two methods,
\textit{redshift-binned} and \textit{Gaussian Process-based}
reconstructions to investigate the effect of a NCC on the equation
of state (EOS) $w(z)$ of evolving dark energy (DE) component. We
find that a NCC is slightly preferred in both the two
reconstructions by up to $\simeq1\sigma$. Although the degeneracy
between the EOS of evolving DE component and NCC weakens the
constraint on the reconstructed $w(z)$, this degeneracy leads to
the phantom divide $w=-1$ more consistent with the 1$\sigma$
posterior of $w(z)$.
    \end{abstract}

    \maketitle

\section{Introduction}
The $\Lambda$CDM concordance model provides the simplest viable
description of our universe and naturally accounts for most
cosmological and astrophysical observations. In this framework,
the cosmological constant (CC), often interpreted as dark energy
(DE), drives the accelerated expansion of the late-time universe.
For decades, determining the nature of dark energy has remained a
key challenge in cosmology (see, e.g., reviews in
\cite{Carroll:2000fy,Frieman:2008sn,Nojiri:2017ncd,Huterer:2017buf}).
While DE is commonly assumed to be a positive cosmological
constant (e.g., \cite{Weinberg:1988cp,Padmanabhan:2002ji}), recent
results from the DESI collaboration \cite{DESI:2025zgx}, combining
baryon acoustic oscillation (BAO) data with Planck cosmic
microwave background (CMB) and Type Ia supernovae (SNe Ia)
measurements, show a statistical preference ($2.8-4.2\sigma$) for
an evolving dark energy model over $\Lambda$CDM when using a CPL
parameterization of the equation of state (EOS)
\cite{Chevallier:2000qy,Linder:2002et} (see also
\cite{DESI:2024lzq,DESI:2024mwx,DESI:2024uvr,DESI:2024aqx,DESI:2024kob}).
This finding has stimulated extensive recent investigation into
dynamical DE scenarios
\cite{Wang:2024dka,Yang:2024kdo,Wang:2024pui,Luongo:2024fww,Cortes:2024lgw,Carloni:2024zpl,Colgain:2024xqj,Giare:2024smz,Gomez-Valent:2024tdb,Escamilla-Rivera:2024sae,Park:2024jns,Shlivko:2024llw,Dinda:2024kjf,Seto:2024cgo,Bhattacharya:2024hep,Wang:2025djw,Wolf:2023uno,Pang:2024qyh,Akarsu:2025gwi,Wolf:2025acj,Roy:2024kni,Heckman:2024apk,Gialamas:2024lyw,Orchard:2024bve,Colgain:2024ksa,Wang:2024sgo,Li:2024qso,Ye:2024ywg,Giare:2024gpk,Dinda:2024ktd,Jiang:2024viw,Alfano:2024jqn,Sharma:2024mtq,Ghosh:2024kyd,Reboucas:2024smm,Wolf:2024eph,RoyChoudhury:2024wri,Arjona:2024dsr,Wolf:2024stt,Giare:2024ocw,Wang:2024tjd,Alestas:2024eic,Carloni:2024rrk,Bhattacharya:2024kxp,Specogna:2024euz,Li:2024qus,Ye:2024zpk,Pang:2024wul,Akthar:2024tua,Colgain:2024mtg,daCosta:2024grm,Chan-GyungPark:2025cri,Sabogal:2025mkp,Du:2025iow,Ferrari:2025egk,Jiang:2025ylr,Peng:2025nez,Jiang:2025hco,Feng:2025mlo,Hossain:2025grx,Chakraborty:2025syu,Borghetto:2025jrk,Pan:2025psn,Pang:2025lvh,Wang:2025ljj,Kessler:2025kju,Yang:2025mws,Wolf:2025jed,RoyChoudhury:2025dhe,Specogna:2025guo,Ye:2025ark,Cheng:2025lod,Ling:2025lmw,Li:2025eqh,Cline:2025sbt,Gialamas:2025pwv,Li:2025dwz,Lee:2025pzo,Ishak:2025cay,Wang:2025vtw,Wang:2025znm,Zhou:2025nkb,RoyChoudhury:2025iis,Pedrotti:2025ccw}.

It is well-known that the anti-de Sitter (AdS) vacuum,
corresponding to a negative cosmological constant(NCC), is both
theoretically significant and well-motivated. Although de Sitter
(dS) vacua can be constructed in principle
\cite{Kachru:2003aw,Kallosh:2004yh}, realizing them within the
string landscape remains nontrivial, particularly in light of
recent swampland conjectures \cite{Ooguri:2006in,Obied:2018sgi}
(see also recent reviews in \cite{Palti:2019pca,Grana:2021zvf} and
\cite{Kallosh:2019axr}). In contrast, AdS vacua arise naturally in
string theory constructions. Recent research has explored the
cosmological implications of anti-de Sitter (AdS) vacua. One
significant focus concerns their potential role in addressing the
Hubble tension, which may be linked to an AdS phase around
recombination
\cite{Ye:2020btb,Ye:2020oix,Jiang:2021bab,Ye:2021iwa,Wang:2022jpo}
or to models with a sign-switching cosmological constant
\cite{Akarsu:2019hmw,Akarsu:2021fol,Akarsu:2022typ,Akarsu:2023mfb,Paraskevas:2024ytz,Anchordoqui:2023woo}.
Concurrently, inflationary evolution within the AdS landscape and
their effects on primordial perturbations have been investigated
\cite{Li:2019ipk,Ye:2022efx,Lin:2022ygd,Piao:2005ag,Piao:2004hr}.

Although a negative cosmological constant cannot itself drive
cosmic acceleration, it can coexist with an evolving, positive
dark energy component. The phenomenological consequences of such
scenarios have been examined combined with CMB and DESI
data\cite{Wang:2024hwd,Notari:2024rti,Mukherjee:2025myk}, and NCC
is preferred by $\simeq1\sigma$, see
\cite{Dutta:2018vmq,Visinelli:2019qqu,Ruchika:2020avj,Calderon:2020hoc,Sen:2021wld,Malekjani:2023ple,Gomez-Valent:2023uof}
for related analysis with pre-DESI baryon acoustic oscillation
(BAO) data and
\cite{Adil:2023ara,Menci:2024rbq,Chakraborty:2025yuo} for analysis
with the James Webb Space Telescope (JWST). See
\cite{Wang:2025dtk} for the scenario with one AdS vacuum around
the recombination and one at low redshift.

Given the unknown essence of dark energy remains, a
non-parametric, data-driven approach can provide a robust way to
probe the evolution of $w(z)$. \textbf{Gaussian process} (GP) has
been widely used in cosmology to reconstruct cosmological
quantities in a model-independent
way\cite{Holsclaw:2010nb,Holsclaw:2010sk,Holsclaw:2011wi,Hwang:2022hla}.
The EOS of dark energy $w(z)$ has been widely reconstructed in
redshift bins without imposing an explicit functional
form\cite{Zhao:2012aw,Zhao:2017cud,DESI:2025wyn,Abedin:2025yru,Wang:2025xvi,Ruchika:2025mkx,Jiang:2024xnu},
see
\cite{Zhang:2018gjb,Elizalde:2018dvw,Gerardi:2019obr,LHuillier:2019imn,Cai:2019bdh,Liao:2019qoc,Aljaf:2020eqh,Benisty:2020kdt,Briffa:2020qli,Li:2025ula}
for other cosmological applications. Although independent from the
pre-recombination physics, these reconstructions assume the
late-time universe dominated by matter ($w_m=0$) and evolving DE
(depicted by $w(z)$).

In this work, we consider an additional CC component
($w_\Lambda=-1$) besides matter and evolving DE component in the
late-time universe, and use two methods, \textit{redshift-binned}
and \textit{Gaussian Process-based} reconstructions to investigate
the effect of NCC on the evolution of $w(z)$. The latter allows us
to provide model-independent insight into the DESI hints for NCC,
as well as the role of the external SNe datasets.

\section{METHODOLOGY}
Considering that dark energy consists of an evolving component and
a cosmological constant simultaneously, the expansion rate is \be
H(z)=H_0\left[\Omega_{m}(1+z)^3+\Omega_{\Lambda}+(1-\Omega_{m}-\Omega_{\Lambda})\exp\left[-3\int_0^z[1+w(z^\prime)]\frac{dz^\prime}{1+z^\prime}\right]\right]^{1/2}\label{H}\ee
where $\Omega_m$ and $\Omega_\Lambda$ are the density parameters
of present-time matter and CC. In this work, we reconstruct the
EOS $w(z)$ of the evolving part of DE, whose density parameter is
$\Omega_x\equiv1-\Omega_{m}-\Omega_{\Lambda}$.

In this work, for all the construction of $w(z)$ we combine the
$\mathbf{DESI}$ DR2 BAO data\cite{DESI:2025zgx} and the
uncalibrated Type Ia SN from the $\mathbf{Pantheon}$
dataset\cite{Scolnic:2021amr}.

\begin{itemize}
    \item [$\bullet$] \textbf{DESI}
    The measurements of DESI for the comoving distance $D_M/r_d$
    and $D_H/r_d$ \cite{DESI:2025zgx},
    \begin{equation}
        D_M(z)\equiv\int_{0}^{z}{c dz'\over H(z')},\quad D_H(z)\equiv {c\over
            H(z)},
    \end{equation}
    where $r_d=\int_{z_d}^{\infty}{c_s(z)\over H(z)}$ is the sound
    horizon, $z_d\simeq1060$ is the redshift at the baryon drag epoch
    and $c_s$ is the speed of sound, as well as the angle-averaged
    quantity $D_V/r_d$ where $D_V(z)\equiv\left(zD_M(z)^2D_H(z)\right)^{1/3}$. The DESI DR2 BAO data cover $0.1<z<4.2$ using multiple tracers: the bright galaxy sample ($0.1<z<0.4$), luminous red galaxies ($0.4<z<1.1$), emission line galaxies ($1.1<z<1.6$), quasars ($0.8<z<2.1$) and Ly$\alpha$ forest ($1.77<z<4.16$).
    \item [$\bullet$] \textbf{Pantheon} We use Pantheon consisting
    of 1701 light curves of 1550 spectroscopically confirmed type Ia
    SN coming from 18 different surveys covering $0.001<z<2.26$\cite{Scolnic:2021amr}.
\end{itemize}

We explicitly implemented all of the components involved in the
DESI BAO and SNe Ia measurements, the likelihood functions can be
written as
\be\mathcal{L}_\mathrm{DESI}=\prod_{i}^{N}\exp\left[-\frac{1}{2}\left(\frac{f(x^i)-y^i}{\sigma_i}\right)\right]\ee
\be\mathcal{L}_\mathrm{Pantheon}=\exp\left[-\frac{1}{2}\Delta
D^TC^{-1}\Delta D\right]\ee where ($x^i,y^i$) and $\sigma^i$ are
the DESI BAO measurements and their uncertainties, $D$ is the
vector of SNe Ia distance modulus residuals and $C$ denotes the
covariance matrices.

We use two methods, \textit{redshift-binned} and \textit{Gaussian
Process-based} reconstructions to investigate the effect of NCC on
the evolution of $w(z)$.
\begin{itemize}
    \item [$\bullet$] \textit{Redshift-binned reconstruction}
We parameterize the equation of state of evolving DE with 10
redshift bins at equal intervals in the range $0<z<2.5$,
characterized by a constant equation of state ($w_i$,
$i=0,1,...,9$) in each bin $0.25i<z<0.25(i+1)$, which is denoted
by $w_\mathrm{bin}$CDM model. Following DESI
paper\cite{DESI:2025wyn}, we use the Big Bang Nucleosynthesis
(BBN) prior
$\Omega_bh^2=0.02196\pm0.00063$\cite{Schoneberg:2024ifp} and CMB
distance information from Planck PR4
$100\theta_s=1.04098\pm0.00042$\cite{Rosenberg:2022sdy} to tighten
the constraint in this redshift-binned reconstruction. \item
[$\bullet$] \textit{Gaussian Process-based reconstruction} To
reconstruct smooth and continuous functions of $w(z)$ and $H_0$
without assuming any explicit functional form, we employ GP as a
non-parametric statistical method. GP assumes any finite
collection of points of a given function $f(x)$, i.e. $w(z)$ in
this work, follows a multivariate Gaussian distribution controlled
by the prior mean $\mu(x)$ and covariance  $k(x,\tilde{x})$. The
$k(x,\tilde{x})$ is typically referred to as kernel, and controls
the strength of correlations between the values of the
reconstructed function at different points, as well as the
deviations from the mean at any given point. For prior
assumptions, we adopt two types of kernels to construct the
covariance matrices, the Gaussian squared exponential kernel \be
k(x_i,x_j)=\sigma_f^2\exp\left(-\frac{(x_i-x_j)^2}{l_f^2}\right)\ee
and the Matern kernel\cite{Holsclaw:2010sk,Seikel:2012uu} \be
k_\nu(r)=\sigma_f^2\frac{2^{1-\nu}}{\Gamma(\nu)}\left(\frac{\sqrt{2\nu}r}{l_f}\right)^\nu
K_\nu\left(\frac{\sqrt{2\nu}r}{l_f}\right)\ee where
$r\equiv|x_i-x_j|$ and $\nu=7/2$ in this work. Without breaking
the degeneracy between $H_0$ and $r_d$, our parameter space
consists of three cosmological parameters
\{$H_0r_d,\Omega_m,\Omega_\Lambda$\} and two kernel hyper
parameters \{$\sigma_f,l_f$\}. To encompasses all the
observational datasets used in this analysis, we model the
evolution of $w(z)$ over the redshift range $0<z<2.5$. Following
\cite{DESI:2025wyn}, the target range is divided into 29 bins of
uniform width in scale factor space.
\end{itemize}
We perform the \textit{redshift-binned} and \textit{Gaussian
Process-based} reconstruction with Markov Chain Monte Carlo (MCMC)
analysis using $\mathbf{Cobaya}$\cite{Torrado:2020dgo}.

\section{Results}
\subsection{Redshift-binned reconstruction}
\begin{figure*}
    \includegraphics[width=0.9\columnwidth]{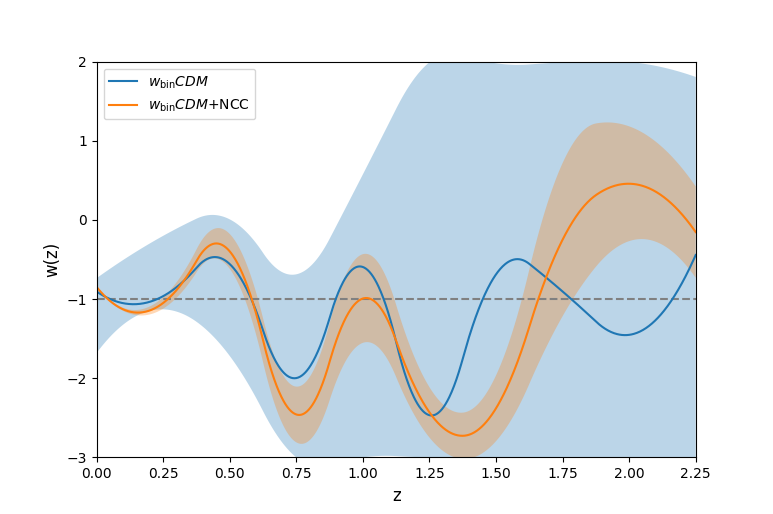}
    \caption{\label{w0}Reconstructed evolution of the redshift-binned dark energy equation-of-state parameter $w(z)$ from DESI+Pantheon data, with and without NCC respectively.}
\end{figure*}

\begin{figure*}
    \includegraphics[width=1\columnwidth]{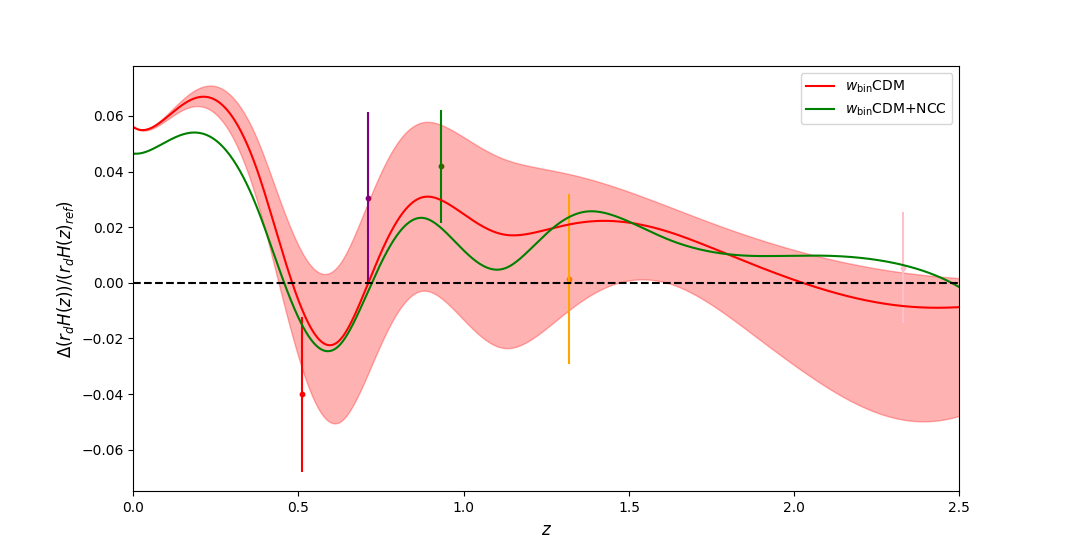}
    \caption{\label{rH}Reconstructed posteriors of the redshift-binned dark energy for BAO observables $r_dH(z)$ with and without NCC respectively. The reference model is the bestfit values of $w_0w_a$CDM in \cite{Wang:2024hwd}.}
\end{figure*}

We construct $w(z)$ of the redshift-binned DE which is displayed
in Fig.\ref{w0}. Marginalized constraints on the cosmological
parameters for $w_\mathrm{bin}$CDM and $w_\mathrm{bin}$CDM+NCC
models are listed in Table \ref{tab1}. A NCC is slightly preferred
in this redshift-binned reconstruction with
$\Omega_\Lambda=-0.093^{+0.525}_{-0.294}$. In light of
insufficient data in $1<z<2$, the EOS of dark energy in this range
is poorly constrained with NCC included, as displayed in
Fig.\ref{w0} and Table \ref{tab1}. Here, from the point of
reconstruction, we find the phantom divide $w=-1$ is consistent
with 1$\sigma$ posterior of $w(z)$ with NCC, which makes the
evolving part of DE quintessence-like. To illustrate how a NCC
component improves the fit to data, we display the BAO observables
$r_dH(z)$ in Fig.\ref{rH}.
\begin{table*}[htbp]
    \centering
    \begin{tabular}{|c|c|c|}
        \hline
        Model&$w_\mathrm{bin}$CDM&$w_\mathrm{bin}$CDM+NCC\\
        \hline
        $w_0$&$-0.847\pm0.005$&$-0.904\pm0.227$\\
        $w_1$&$-1.048\pm0.072$&$-0.962\pm0.225$\\
        $w_2$&$-0.397\pm0.220$&$-0.557\pm0.315$\\
        $w_3$&$<-2.224$&$<-0.686$\\
        $w_4$&$-0.992\pm0.560$&$<0.576$\\
        $w_5$&$-2.430\pm0.318$&Unconstrained\\
        $w_6$&$<-2.154$&Unconstrained\\
        $w_7$&$-0.249\pm0.967$&Unconstrained\\
        $w_8$&$>0.089$&Unconstrained\\
        $w_9$&$-0.153\pm0.577$&$<1.807$\\
        \hline
        $H_0$&$67.81\pm0.53$&$67.99\pm0.79$\\
        $\Omega_m$&$0.321\pm0.005$&$0.305\pm0.008$\\
        $\Omega_\Lambda$&-&$-0.093^{+0.525}_{-0.294}$\\
        \hline
    \end{tabular}
    \caption{\label{tab1}Mean and $1\sigma$ values for $w_\mathrm{bin}$CDM and $w_\mathrm{bin}$CDM+NCC models with DESI+Pantheon dataset.}
\end{table*}

\subsection{Gaussian Process Regression}
\begin{figure*}
    \includegraphics[width=0.45\columnwidth]{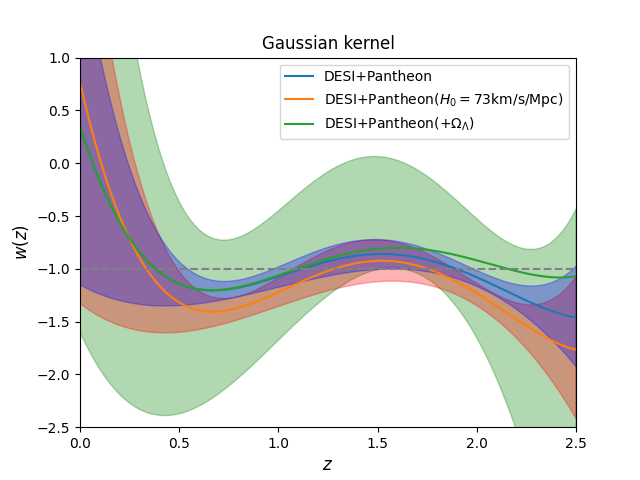}
    \includegraphics[width=0.45\columnwidth]{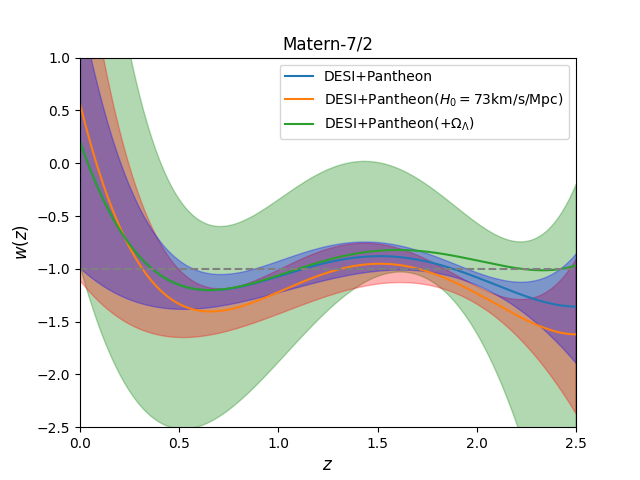}
    \caption{\label{w}Reconstructed evolution of the dark energy equation-of-state parameter $w(z)$ from DESI+Pantheon data, as well as that with fixed $H_0$ and extra $\Omega_\Lambda$ respectively.}
\end{figure*}

\begin{table*}[htbp]
    \centering
    \begin{tabular}{|c|c|ccc|}
        \hline
        kernel&model&$H_0r_d$&$\Omega_m$&$\Omega_\Lambda$\\
        \hline
        \multirow{3}{*}{Matern}&-&$100.460\pm0.831$&$0.314\pm.008$&-\\
        \cline{2-5}
        &fixed $H_0$&-&$0.298\pm.007$&-\\
        \cline{2-5}
        &$+\Omega_\Lambda$&$103.271\pm0.831$&$0.305\pm.010$&$-0.729^{+0.797}_{-1.137}$\\
        \hline
        \multirow{3}{*}{Gaussian}&-&$100.730\pm0.805$&$0.314\pm.008$&-\\
        \cline{2-5}
        &fixed $H_0$&-&$0.299\pm.007$&-\\
        \cline{2-5}
        &$+\Omega_\Lambda$&$102.943\pm0.831$&$0.303\pm.008$&$-0.613^{+0.805}_{-1.012}$\\
        \hline
    \end{tabular}
    \caption{\label{tab}Marginalized constraints on the cosmological parameters for various kernels and models.}
\end{table*}
We construct $w(z)$ of DE and using two distinct kernel functions,
Matern-7/2 and Gaussian kernels for comparison, which are
displayed in Fig.\ref{w}. Marginalized constraints on the
cosmological parameters for various kernels and models are listed
in Table \ref{tab}. Constructions with the two kernels give
similar results, and support a NCC by $\simeq1\sigma$. The
reconstructed $w(z)$ crosses into the phantom regime ($w<-1$)
around $z\simeq0.5$ in all the cases, consistent with previous
non-parametric DESI BAO analysis\cite{Wang:2025xvi,DESI:2025wyn}.
When the existence of NCC is included, several deviations from
evolving DE appear at distinct epochs. We find the best-fit $w(z)$
is more consistent with $w=-1$ and the viable difference from the
case without CC appears only in $z\gtrsim1$. However, the
evolution is weakly constrained when CC is included, which
attributes to the additional parameter $\Omega_\Lambda$. The
constraint on $\Omega_\Lambda$ is weak as a result of the
degeneracy between $\Omega_\Lambda$ and $w(z)$. This aligns with
our previous MCMC result of CPL-like DE and CC that NCC increases
the density of evolving DE and shrinks the evolving behavior to
reproduce the CPL-like result\cite{Wang:2024hwd}. Considering that
NCC potentially helps to lift
$H_0$\cite{Adil:2023ara,Menci:2024rbq,Chakraborty:2025yuo}, we
also display the construction of $w(z)$ with $H_0$ fixed to
$H_0=73\mathrm{km/s/Mpc}$ and approximated $r_d$ according to
Ref.\cite{DESI:2025zgx}. When a high $H_0$ is fixed, a smaller
$w(z)$ in $z\gtrsim0.5$ is preferred, consistent with our previous
MCMC results considering the $H_0$ prior of SH0ES and CPL-like
DE\cite{Wang:2024dka}.

\begin{figure*}
    \includegraphics[width=0.7\columnwidth]{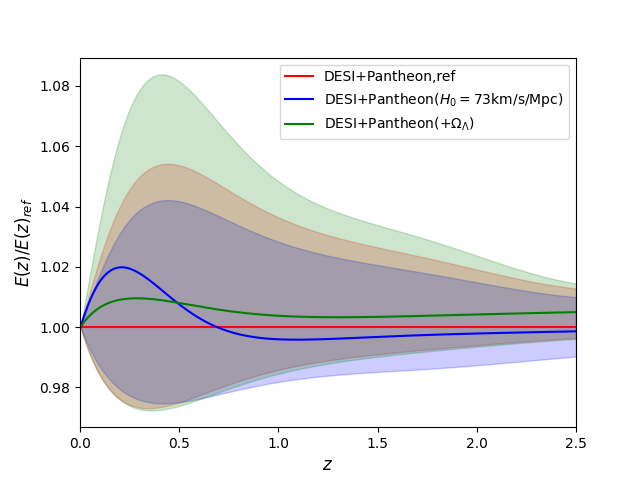}
    \caption{\label{E}Reconstructed posterior using squared exponential kernel for the expansion rate $E(z)$ relative to the reference model.}
\end{figure*}

According to (\ref{H}) NCC alters the late-time evolution which
can be reproduced by the reconstructed $w(z)$. Our non-parametric
reconstruction of the expansion rate $E(z)\equiv H(z)/H_0$ is
shown in Fig.\ref{E}, which indicates that deviations are limited
to $\lesssim10\%$ relative to the reference evovling DE
reconstruction. Both NCC and a high $H_0$ lead to a larger $E(z)$
in the range $z<0.5$, while their effect differs at $z>1$. In
light of the distinct effects on $w(z)$ caused by $\Omega_\Lambda$
and a high $H_0$, NCC could not resolve Hubble tension merely
considering the modification of late-time evolution.

\begin{figure*}
    \includegraphics[width=1\columnwidth]{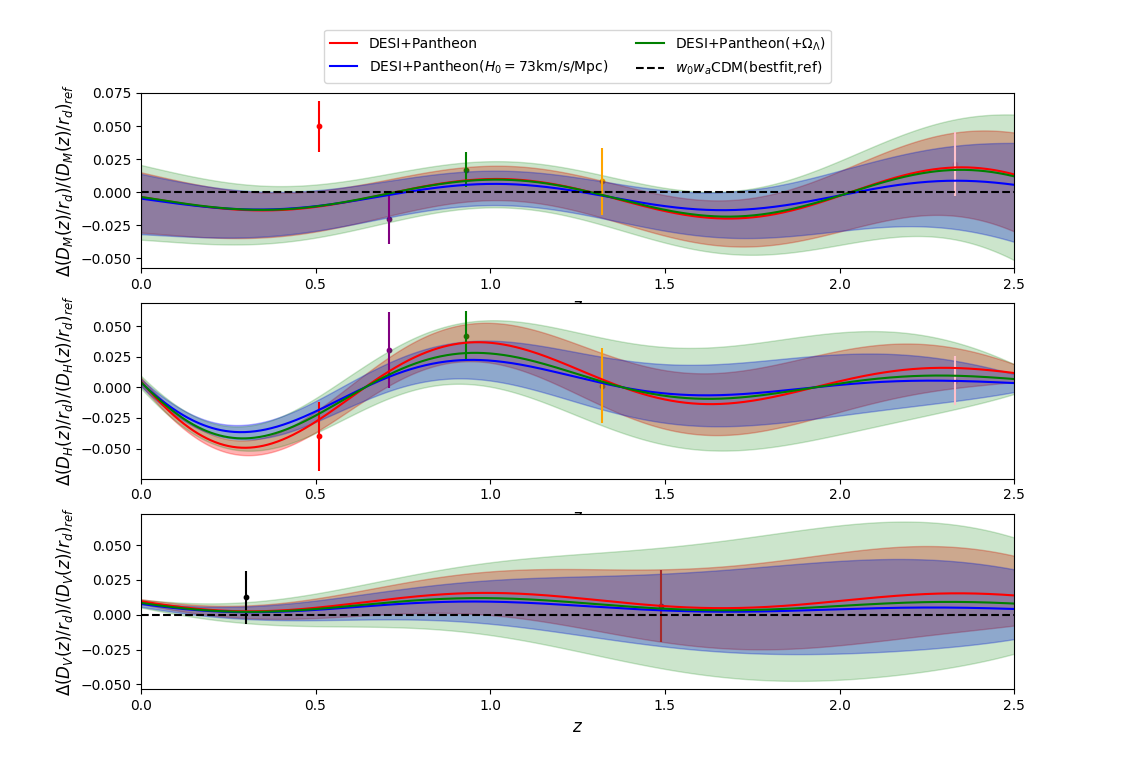}
    \caption{\label{D}Reconstructed posteriors using squared exponential kernel for BAO observables $D_M(z)/r_d$, $D_H(z)/r_d$ and $D_V(z)/r_d$ for different models respectively.}
\end{figure*}

To illustrate how the data constrains the late-time evolution
$w(z)$, we display the reconstructed $D(z)/r_d$ related to the
best-fit values of $w_0w_a$CDM model. All reconstructions show
better fitting especially to $D_H(z)/r_d$ points. Considering a
NCC, the evolution at $z<0.5$ is extremely consistent with the
evolving DE construction. The evident discrepancy of 1$\sigma$
region is only viable at higher red shift $z\gtrsim1$, while the
shapes are still well consistent. This explains why we have
similar $\chi^2$ in MCMC results\cite{Wang:2024hwd}. In
\cite{Wang:2024hwd}, we sample
\{$w_{0,\mathrm{eff}},w_{a,\mathrm{eff}}$\} using MCMC rather than
\{$w_0,w_a$\} of CPL-like DE to capture the constraint on
$\Omega_\Lambda$. This attributes to the weak effect of
$\Omega_\Lambda$ on the zeroth and first orders of the evolution
depicted by \{$w_{0,\mathrm{eff}},w_{a,\mathrm{eff}}$\}, and
merely the second order is much influenced by $\Omega_\Lambda$ as
we see in Fig.\ref{D}. Therefore our selection of parameters avoid
potential volume effect caused by the degeneracy between
\{$w_{0},w_{a}$\} and $\Omega_\Lambda$.

\section{Discussion}
In this paper, we investigate the effect of a NCC $\Omega_\Lambda$
on the late-time evolution using parametric and non-parametric
reconstruction methods. We reconstruct the evolving DE EOS $w(z)$
together with a NCC $\Omega_\Lambda$ by redshift-binned dark
energy and a non-parametric method, i.e. the Gaussian process,
with the latest DESI BAO and SNe data. We find that a NCC is
slightly preferred by up to $\simeq1\sigma$ in both
redshift-binned and Gaussian process-based reconstruction, and the
bestfit reconstructed evolution of $w(z)$ is roughly consistent
with the case without NCC at $z\lesssim1$, with deviations that
could be captured by observations only displayed at higher red
shift $z\gtrsim1$. This insensitivity of $\Omega_\Lambda$ on
observables at low red shift explains the degeneracy between
\{$w_0,w_a$\} in CPL-like DE and $\Omega_\Lambda$ using MCMC
analysis, as pointed out in the
literature\cite{Wang:2024hwd,Notari:2024rti}. From the perspective
of reconstruction, this degeneracy between the EOS of DE and NCC
also leads to the phantom divide $w=-1$ consistent with the
1$\sigma$ posterior of $w(z)$ with NCC, which makes
quintessence-like DE possibly reliable.

Compared with the $w(z)$ reconstruction with a fixed
$H_0=73$km/s/Mpc, the discrepancy indicates the poor effect on
resolving Hubble tension by merely introducing NCC. However, NCC
could still help to alleviate Hubble tension combined with
pre-recombination scenarios, especially early dark energy. In
\cite{Wang:2025dtk}, the model with two anti-de Sitter vacua
existing in both early and late times has been investigated and
shows the potential of NCC to resolve Hubble tension.

Considering DE energy density is sub-dominant $\lesssim10\%$ at
$z\gtrsim2$, we emphasize the importance of the observation at
$1\lesssim z\lesssim2$ on constraining $\Omega_\Lambda$. Although
current observations could not provide tight enough constraint on
NCC, hopefully the further data of DESI and upcoming Euclid
\cite{Euclid:2024yrr} combined might improve our understanding of
the nature of NCC.

\section*{Acknowledgments}
This work is supported by NSFC, No.12475064, National Key Research
and Development Program of China, No.2021YFC2203004, and the
Fundamental Research Funds for the Central Universities.

\end{document}